# Resiliency and Robustness of Complex, Multi-Genre Networks

*Alexander Kott (ARL), Tarek Abdelzaher (UIUC)*

## 1. Introduction to Complex Networks

In this paper, we explore the resiliency and robustness of systems while viewing them as complex, multi-genre networks. We show that this perspective is fruitful and adds to our understanding of fundamental challenges and tradeoffs in robustness and resiliency, as well as potential solutions to the challenges.

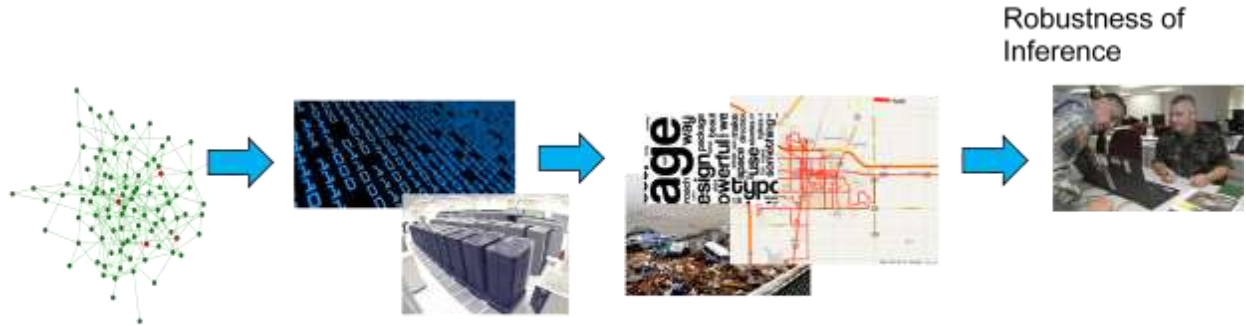

**Figure 1.** A complex network often includes networks of several distinct genres – networks of physical resources, communication networks, information networks, and social and cognitive networks.

When using the term "*complex, multi-genre networks*" we refer to networks that combine several distinct genres – networks of physical resources, communication networks, information networks, and social and cognitive networks. As illustrated in Fig. 1, a complex network may utilize a physical resource network (e.g., a sensor network to measure and report on phenomena of interest), to store information (e.g., a data cloud), and to communicate it to the stakeholders (e.g., the Internet, or a courier network), and a network of human decision makers (and possibly artificial agents) that assess and comprehend the information and produce decisions.

One of the key commodities flowing through such a network are data elements that themselves are connected by links, either implicit (e.g., semantic relations) or explicit (e.g., URLs), forming a web, called an information network, from which reliable information is to be distilled. Other elements and links in the complex network could be of physical nature, such as actuators that execute the decisions made by human or artificial agents, or a power grid, a road network, etc. The combination of these heterogeneous networks forms the network ecosystem whose resiliency we wish to understand. A resilient ecosystem will continue to support sound decisions and actions even as physical infrastructure is damaged, sources are infiltrated, data is corrupted, or access to critical resources is denied.

Study of systems such as multi-genre networks is relatively uncommon; instead, it is customary in research and engineering literature to focus on a view of a network comprised of homogeneous elements, (e.g., a network of communication devices, or a network of social beings). Yet, most if not all real-world networks are multi-genre – it is hard to find any real system of a significant complexity that does not include a combination of interconnected physical elements, communication devices and channels, data collections, and human users forming an integrated, inter-dependent whole.

The multi-genre perspective becomes more important as our society and socio-technical systems are permeated with growing number of communication links between humans and physical systems (e.g., Internet, cell phones) and corresponding social links (e.g., the broad popularity and impact of social networking applications). Note that, when we consider multiple genres of networks, the total number of links that connect any two elements grows significantly. For example, two physical devices may not be connected by any physical or communication links, but if their human owners are part of the same social network, the devices are in fact connected through their human owners, in ways that may be significant for understanding the devices' behavior. Additional links, and especially additional links of heterogeneous nature, are likely to affect the network's robustness and resiliency.

For this reason, our ability to design, manage and operate many industrial, financial, and military systems has become increasingly dependent on our understanding of a complex web of interconnected physical infrastructure, distributed computation, and social elements. It is the combined, interactive behavior of multiple genres of networks that determines the ultimate properties of the overall system, including robustness and resiliency. Further complexity arises because such systems often exist in adversarial environments and in the presence of resource loss, uncertainty, and data corruption.

Making the right decisions, and more generally, exerting the desired influence upon such systems, requires proper modeling of networked physical, information, and social elements to understand their combined behavior, assess their vulnerabilities, and design proper recovery strategies. In recent years, research programs have been initiated to understand the interactions of dissimilar genres of networks (e.g. [32]).

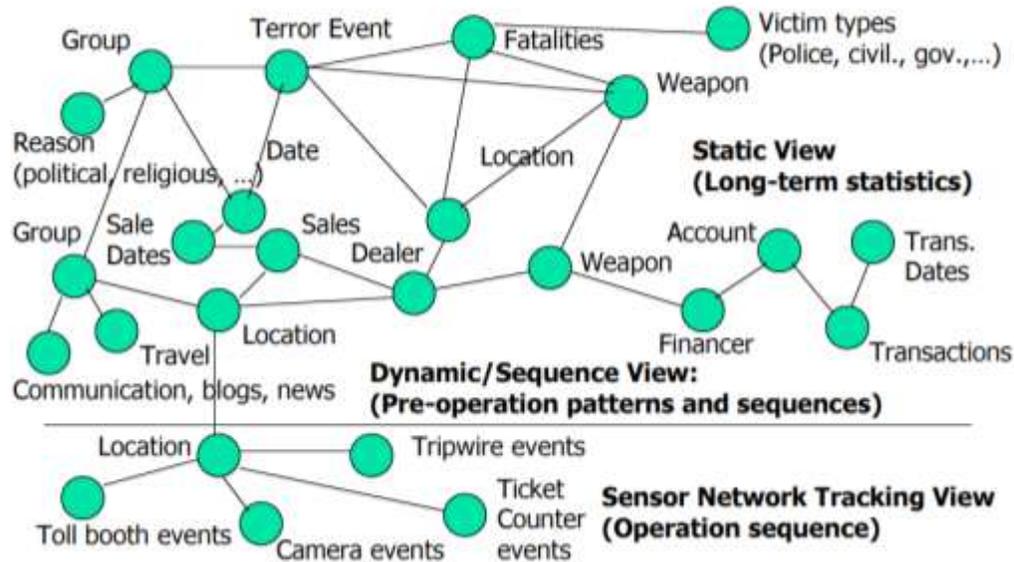

**Figure 2** A counter-terrorism network: This complex network includes nodes and links that belong to multiple genres: information, communications, social.

## 2. Resiliency and Robustness

As discussed earlier in this book, robustness and resiliency are related yet distinct properties of systems. Robustness denotes the degree to which a system is able to withstand an unexpected internal or external event or change without degradation in system's performance. To put it differently, assuming two systems – A and B—of equal performance, the robustness of system A is greater than that of system B if the same unexpected impact on both systems leaves system A with greater performance than system B.

We stress the word *unexpected* because the concept of robustness focuses specifically on performance not only under ordinary, anticipated conditions (which a well-designed system should be prepared to withstand) but also under unusual conditions that stress its designers' assumptions. For example, in IEEE Standard 610.12.1990, "Robustness is defined as the degree to which a system operates correctly in the presence of exceptional inputs or stressful environmental conditions." Similarly, "robust control refers to the control of unknown plants with unknown dynamics subject to unknown disturbances" [25].

The resiliency, on the other hand, refers to the system's ability to recover or regenerate its performance after an unexpected impact produces a degradation of its performance. For our purposes, assuming two equally performing systems A and B subjected to an unexpected impact that left both systems with an equal performance degradation, the resiliency of system A is greater if after a given period T it recovers to a higher level of performance than that of system B.

For example, ecology researchers describe resiliency as "the capacity of a system to absorb disturbance and reorganize while undergoing change so as to still retain essentially the same function, structure, identity, and feedbacks" [26]. Alternatively, one may focus on the temporal aspect of the definition, where resiliency is "the time required for an ecosystem to return to an equilibrium or steady-state following a perturbation" [27]. Note that, in complex non-linear systems, the length of recovery typically depends on the extent of damage. There may also be a point beyond which recovery is impossible. Hence, there is a relation between robustness (which determines how much damage is incurred in response to an unexpected disturbance) and resiliency (which determines how quickly the system can recover from such damage). In particular, a system that lacks robustness will often fail beyond recovery, hence offering little resiliency. Both robustness and resiliency, therefore, must be understood together.

Does the multi-genre network perspective offer anything to our understanding of robustness and resiliency? It does, because it changes our perspective on complexity of the links within the system. In his pioneering work, Perrow [28] explains that catastrophic failures of systems emerge from high complexity of links which lead to interactions that the system's designer cannot anticipate and guard against. System's safety precautions can be defeated by hidden paths, incomprehensible to the designer because the links are so numerous, heterogeneous, and often implicit. Greater connectivity we recognize in a multi-genre network helps us see more of the overall network's complexity, and hence the potential influences on its robustness and resiliency.

For example, greater complexity of connections between two elements of the systems, such as the number of paths that connect them through links of various natures, may lead to increased robustness of the system by increasing redundancy of its functions. The same increase in complexity, however, may also lead to lower robustness by increasing—and hiding from the designer—the number of ways in which one failed component may cause the failure of another. When we consider an entire multi-genre network— and not merely one of the heterogeneous, single-genre sub-networks that comprise the whole – we see far more complexity of the paths connecting the network's elements.

A system of military command – a multi-genre network that comprises human decision-makers, sensing, communication and computing devices and large collections of complex interlinked information— exhibits complex modes of failures that become more likely as the degree of collaboration between the system's elements increases. Experimental studies of such systems [29] show how increased availability of information delivered through an extensive communications network may paralyze decision makers and induce them into a confusion or endless search for additional information. They also show how increased collaboration—made possible by improved networking—may mislead collaborators into accepting erroneous interpretations of the available information.

Of particular importance are those paths within the system that are not recognized or comprehended by the designer. Indeed, the designer can usually devise a mechanism to prevent a propagation of failure through the links that are obvious. Many, however are not obvious, either because there are simply too many paths to consider – and the numbers rapidly increase once we realize that the paths between elements of a communication system, for example, may also pass through a social or an information network—or because the links are implicit and subtle. Kott [30] offers examples where subtle feedback links lead to a failure in organizational decision making. It is even more difficult to comprehend an enormous number of complex, multi-genre links in large scale societal effects, such as insurgency dynamics [31].

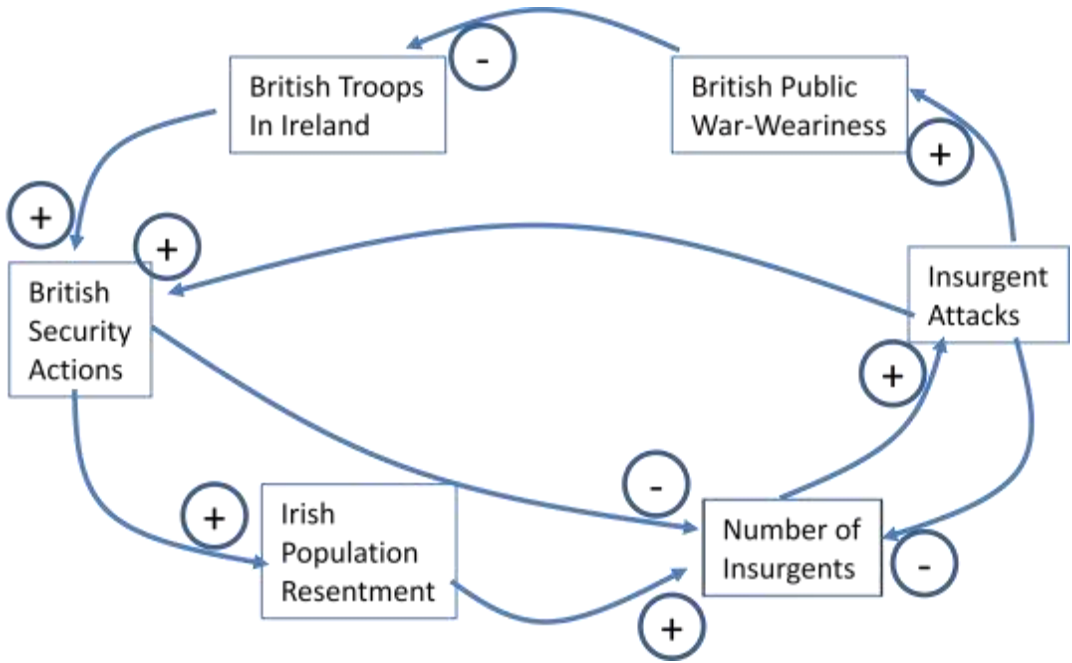

**Figure 3.** Even in this highly simplified network of insurgency dynamics in Northern Ireland there are multiple, often non-obvious paths between phenomena in the system [31].

Similarly, greater complexity of a network may have a range of influences on resiliency. As the number and heterogeneity of links grow, they offer the agents (or other active mechanisms) within the network more opportunities to regenerate the network's performance. These agent may be able to use additional links to more elements in order to reconnect the network, to find replacement resources, and ultimately to restore its functions. This does however require that the network includes the agents capable of taking advantage of the opportunity offered by increased complexity. On the other hand, greater complexity of the network may also reduce the resiliency of the network. For example, agents may be more likely to be confused by the complexity of the network, or to be defeated in their restoration work by un-anticipated side effects induced by hidden paths within the network. In either case, by considering the holistic, multi-genre nature of the network, the designers or operators of the system have a better chance to assess its robustness and resiliency.

### 3. Fundamental Tradeoffs

Most approaches to improving resiliency and robustness involve compromises, and the key challenge is to find a favorable compromise [33]. Such compromises involve reducing or managing the complexity of the network: coupling, rigidity and dependency. We discuss several of these approaches in the ensuing sections of this paper:

*Resources vs resiliency*: additional resources in a network can help improve resiliency. For example, adding capacity to nodes in a power generation and distribution network may reduce likelihood of cascading failures and speed up the service restoration (as discussed in Section 4 and Section 5). Adding local storage and influencing the distribution of nodes of different functions in a supply network also leads to improved resilience at the expense of additional resources (Section 6).

*Performance vs resiliency*: a network optimized for higher performance may experience greater degradation under an impact, and it may be more difficult and time consuming to restore its operation (Section 4). For example, it was shown that "sub-optimal" design that is risk-averse will generally avoid disastrous failure with a higher probability than the optimal one [2]. A power generation and distribution network controlled by a sophisticated computer network will yield better performance, yet may experience greater and harder-to-restore cascading collapse (Section 5).

*Resiliency to one type of disruption vs resiliency to another disruption type*: in order to improve a network's resiliency against a certain type of impact, designers may have to sacrifice its resiliency against another type of impact. For example, in a scale-free network a targeted attack may produce much greater damage (and is more difficult to rectify) than a random attack or random failure (Section 4, Section 5).

*Complexity vs resiliency*: improved resiliency can be obtained by incorporating more sophisticated mechanisms into a network. Complexity goes beyond merely adding resources to the system. Complexity implies potentially a more costly system, and most likely greater investment into development, operation and maintenance of the system. For example, resiliency may be improved by adding multiple functional capacities to each node (Section 6), by processing more input sources (Section 7), or by combination of multiple inference algorithms (Section 8). Yet the same complex measures might cause greater difficulties in restoring the network's capability if degraded by an unexpected -- and probably harder to understand-- failure.

Several tradeoffs often coexist and amplify the designer's challenge.

Finally, it is important to notice that in multi-genre networks, there is an inseparable relation between resiliency and robustness in that the latter is a precondition of the former. This relation comes from the interactions that such systems have with their physical and social environment. To explain, observe that when we have a system with low robustness (e.g., small failures cascade to catastrophes) that is partly "physical", it is hard to build resiliency on top. This is because physical failures and human losses are often irreversible or expensive to remedy (e.g., once there is a reactor meltdown, it is hard to "roll back"). This characteristic is unlike purely logical systems (e.g., databases), where roll-back from failure is cheaper, making robustness and resiliency somewhat orthogonal concerns.

## 4. Robustness to Topology Modifications

The first precondition of resiliency in multi-genre networks is robustness to topology modifications. This robustness allows the network to survive local damage (that results in topological change). Most biological beings, for example, are extremely resilient in that they are able to recover remarkably well from local injury or damage. However, resiliency is not limitless. There are bounds on injury or illness within which resiliency mechanisms work well. If damage exceeds those bounds, recovery in general cannot be attained. Hence, in order to engineer resilient systems, it is important to understand their underlying robustness properties first, since these properties help understand the bounds within which recovery (and hence resiliency) is possible. Systems with low robustness will have a very limited range in which recovery is efficient. The infrastructure on which multi-genre networks rest consists of multiple inter-dependent physical networks, such as a road network, a data network, or the power grid. The first key characteristic that determines response to failure of such networks is their topology. It is therefore interesting to ask the question: in a system described by an arbitrary graph or interconnected network, what topologies are inherently more robust?

Much prior research addressed the fundamental vulnerabilities of different networks as a function of their topological properties. Of particular interest has been the classification of properties of networks according to their node degree distribution. While some networks (such as wireless and mesh networks) are fairly homogeneous and follow an exponential node degree distribution, others, called scale-free networks (such as the Web or the power-grid), offer significant skew in node degrees, described by a power-law. In a key result by Albert, Jeong, and Barabási [3], it was shown that scale-free graphs are much more robust to random node failures (errors) than graphs with an exponential degree distribution, but that these scale-free graphs are increasingly more vulnerable to targeted attacks (namely, removal of high-degree nodes). The above observations are intuitive and can be explained by the difference in topology between scale-free and exponential networks.

An exponential node degree distribution means that nodes with a larger degree are exponentially less probable. This makes the network more homogeneous. In contrast, in a scale-free network, node degree is distributed according to a power law. The power law distribution has a heavy tail, which means that some nodes are extremely well connected, whereas most nodes are not. The aforementioned difference in the degree of homogeneity versus skew in the two types of networks explains the difference in their robustness properties. Specifically, homogeneous networks, by the very nature of their homogeneity, feature uniform degradation as nodes are removed, and offer no significant difference in behavior when the most connected nodes are removed first. This is because all nodes contribute roughly equally to network connectivity. In contrast, power-law networks, due to their skewed distribution, are less sensitive to random removal, which tends to affect the less connected nodes more (since there is more of them) and hence tends to have a less significant effect on connectivity compared to in homogeneous networks. Targeted removal, on the other hand, causes much more damage by eliminating the relatively few highly connected nodes.

For the same reason, as one might expect, when nodes are randomly removed, the size of the largest remaining connected component in the network is bigger in the case of power-law networks than in the case of exponential networks. However, if highest-degree nodes are removed first, the opposite is true. This has implications on the extent of effort needed, for example, to partition the network, both in the case of an attack and in the case of random failures. Callaway, Newman, Strogatz and Watts [4] describe an analytic technique for quantifying the above trends, by relating them to appropriate percolation problems in random graphs with an arbitrary degree distribution. The underlying basic theory for deriving properties of networks of arbitrary degree distribution is described in [1].

The above results are consistent with the intuition that performance optimizations exacerbate worst-case behavior. Scale-free networks perform better in the sense of remaining connected longer despite random errors, but the very mechanism that allows them to perform better (namely, the existence of a few very well-connected nodes) also causes increased vulnerability in the worst-case scenario. This property is often called "*robust yet fragile*", and is a characteristic of many large-scale complex networks [23].

While we cannot change the topology of a large network, such as the Internet, in order to improve robustness as a foundation for resiliency, the network should have a mechanism to reduce dependencies. Dependencies, in this context, exist because performance depends on the availability, capacity, and latency of the few well-connected nodes. In a later section, we discuss how such dependencies are relaxed. For example, on the WWW, one of the most common power-law networks in use, content is simply replicated (cached). Retrieving the most popular content from a cache reduces dependence on the well connected sources (e.g., data centers) and backbone routers at the expense of some degradation in content freshness.

## 5. Robustness to Cascaded Resource Failures

The next step in understanding robustness as a precondition of resiliency is to understand robustness of a network to cascaded failures. Such failures are non-independent in that one triggers another. A network that is prone to large "domino effects" will likely sustain severe damage in response to even modest disturbances, which significantly limits the scope within which efficient recovery (and hence resilient operation) remains possible. Hence, below, we extend the results presented in the previous section to non-independent failures.

Multi-genre networks exhibit pathways by which component failures may cascade from one node to another; an issue not discussed above. It is of great interest to analyze the susceptibility of networks to cascaded failures, such as large blackouts or congestion collapse. Early mathematical analysis of failure cascades as a function of different network topologies was published by the National Academy of Sciences [8]. A model was assumed, where a node fails if a given *fraction* of its neighbors have failed. The model, although applicable to a class of failure propagation scenarios, was intended to be general enough to describe other cascades as well, such as rumor propagation and propagation of new trends. Consistently with results analyzing independent failures, it was shown that, under this propagation model, scale-free networks are less susceptible to propagation cascades emanating from random failures. This may have been expected given their higher robustness to *independent* random failures as well. Note also, that under this model, since the failure threshold is set on the *fraction* (not the absolute number) of failed neighbors, nodes with a higher degree generally need more neighbors to fail first before they collapse as well, which mitigates propagation of the cascade in scale-free networks.

A failure propagation model that is more relevant to flow networks was described in [12]. In this model, each node is assumed to send a unit of flow to each other node via the shortest path. Hence, the expected load on a node, and thus its capacity, should be proportional to its centrality, or how many paths it falls on. The higher the proportionality factor between expected load and maximum capacity, the more extra leeway exists to accommodate overload. A node is assumed to fail if its actual load exceeds capacity. This may occur if other nodes fail, causing flow re-routing. Hence, this model aims at understanding overload-related failure cascades. Unsurprisingly, it was shown that scale-free networks, according to this model, are more vulnerable to failure cascades due to attacks on *selected* (i.e., high-degree) nodes than homogeneous networks, but are less susceptible to cascades that result from *random* failures.

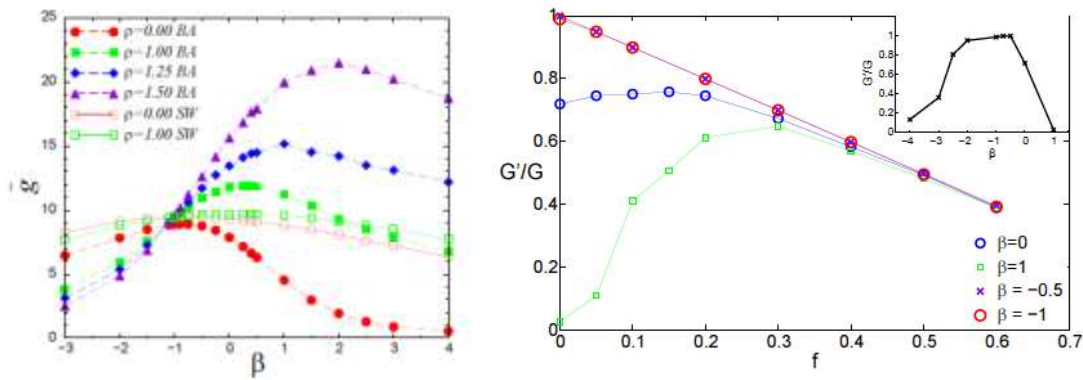

**Figure 4.** Network vulnerability can be minimized by managing the relation between properties of links and node degrees. (Adapted from [13].) Left shows network conductance, g, as a function of edge weight parameter, $\beta$, for different source/target distributions, $\rho$. Right shows the relative size of the surviving component, G'/G, as a function of the fraction of removed nodes, f, for different $\beta$.

Recent work [13] further analyzed the trade-off between network throughput (or conductance) and robustness to failure in scale-free networks with *weighted* links. It was shown that while these networks remain vulnerable to cascades due to targeted destruction of highly-connected nodes, their vulnerability is minimized when link conductance (weight) is set inversely proportional to the product of the linked node degrees.

While the above results appear to be mutually consistent, later evidence showed that the exact nature of failure propagation dramatically affects the robustness properties of networks, often contradicting established intuitions. For example, an interesting analysis of cascaded failures is found in recent work on interdependent networks [6]. It is inspired by a blackout example, where failure of nodes in a power grid causes failure of nodes in a powered computing and communication network, but since some of these computing nodes are used to control the grid, their failure, in turn, causes further failures in the power grid, resulting in a vicious cycle. If the dependencies between the networks are modeled by links, it was shown that the set of interdependent networks was more susceptible to cascaded failures than the individual networks in isolation. Moreover, in an apparent contradiction to prior intuition, an interdependent set of scale-free networks was shown to be *more susceptible to random failures* than an interdependent set of exponential networks. This may appear to be puzzling, considering that scale-free networks are known to be *more robust to random failures* than their homogeneous counterparts. The authors explained that the phenomenon was because, in inter-dependent scale-free networks, low-degree nodes in one network are more likely to propagate failures to high degree nodes in another [6]. Indeed, contrary to prior work, where failure of a node was conditioned on failure of a *fraction* of its neighbors [8], the interdependent-network study assumed that a *single* failed neighbor in the other network brings down a node. Hence, nodes with a higher-degree in one network would tend to die quickly when random failures occur in the other, making scale-free networks (where connectivity depends on such high-degree nodes) increasingly susceptible.

Indeed, much of the results on network robustness in the presence of failure cascades depend on the model of the failure cascade. This dependence on failure propagation models was articulated well in a recent study [7], where it was shown that radically different graphs, such as cliques versus trees, can become more robust under different models of failure propagation. For example, if a node's failure probability *increases with the number of its failed neighbors* (as in the web of interdependent financial establishments), having diversity among neighbors (i.e., lack of "dependency" edges) is preferable, in the sense of minimizing failure risk. This is so regardless of the size of the overall connected component to which the nodes belong. For a given number of edges per node, a tree topology minimizes edges among a node's neighbors and hence minimizes risk in this scenario. In contrast, if the failure of *any one neighbor* is sufficient to bring-down a node (as in the propagation of infectious diseases) then the failure probability is a function of the size of the connected component to which the node belongs, since a domino effect that starts anywhere in the component will eventually bring down all nodes. Given the same number of edges per node, a clique (not a tree) would minimize the size of the connected component and hence minimize risk. By a similar argument, scale-free versus homogeneous networks would also be better for different failure propagation models. The example suggests the importance of modeling failure propagation correctly when analyzing network robustness. Proper modeling of failure propagation is especially challenging in networks where node interactions are mitigated by computing or other intelligent engineered components (e.g., computer-controlled switches in a power grid or on the Internet). The algorithms implemented by such components will likely significantly affect the failure propagation model and hence the resiliency properties of the network at hand. An interesting question therefore becomes to engineer these components in ways that maximize resiliency of the current topology.

## 6. Buffering and Resiliency to Function Loss

The vulnerability of the underlying resource network to connectivity breaches (e.g., as measured by the size of the surviving connected component) is only one factor affecting the overall vulnerability of the entire multi-genre network. In general, the overall performance of the process depends on other factors besides connectivity, such as the degree to which it experiences *function loss* as a result of various failures and perturbations.

In data and commodity flow networks, the function of the network is to offer its clients access to a set of delivered items. In such networks, buffers (e.g., local distributors) constitute a resiliency mechanism that obviates the need for continued access to the original source. Should the original source become unreachable, one can switch to a local supplier. Hence, local access can be ensured despite interruption of the global supply network as long as access to a local distributor (buffer) is available. Local access is an especially valuable solution in the case of a data flow network, where the commodity (namely, the data content) is not consumed by user access, in the sense that a local distributer can continue to serve a content item to new users irrespective of its use by others.

Much work on network buffering has been done to increase the resiliency of data access to fluctuations in resource availability. For example, buffering (or caching) has been used to restore connectivity and performance upon topology changes in ad hoc networks [11], as well as to reduce access latency in disruption-tolerant network [14].

The concept of buffering, and its relation to resiliency, however, transcends data and commodity flow networks in general. In recent work on biological modeling [9], the concept of *buffering* was generalized to model the extent to which degeneracy in multi-functional agents offers "spare capabilities" for adaptation. Degeneracy, a term borrowed from biology, refers to a condition where (i) agents (network nodes) can perform one of multiple functions depending on context, and (ii) the same function can be performed by one of several agents. For example, storage agents (buffers) in a commodity flow network can use their space to store any of a set of possible items. Also, the same item can be stored by any of multiple agents, hence fulfilling the conditions of degeneracy. The degeneracy model applies in other contexts as well. For example, individuals in an organization can allocate their time to any of a set of possible projects. Similarly, the same project can be performed by any of multiple individuals. It is shown that functional degeneracy (i.e., the combination of versatility and redundancy of agents) significantly improves resiliency of network functions [9] by facilitating reconfiguration to adapt to perturbations. Intuitively, the higher the versatility of the individual agents and the higher the degree to which they are interchangeable, the more resilient is the system to perturbation because it can reallocate functions to agents more flexibly to restore its performance upon resource loss. Degeneracy and networked buffering also lead to a better adaptive behavior to changing conditions, as argued, for example, in the context of agile manufacturing [10].

A dimension not typically explored in biology (where agents, such as cells, can perform one main function at a time), is the dimension of *capacity*; that is, the number of different functions that an agent can simultaneously perform. Capacity quantifies, for example, the number of items a storage node can simultaneously hold, or the number of projects a given individual can simultaneously work on. Quantifying the resiliency of multi-genre networks as a function of both their degeneracy and their capacity remains an interesting open problem. The problem is especially interesting in the context of non-independent failures; that is to say, in situations where failures of some nodes or functions can propagate to others, according to some appropriate propagation model.

In the context of data fusion processes, one example of functional loss in networks with degeneracy and non-independent failures comes from distributed real-time computing. Processors (nodes) in such computing systems run several computational workflows with end-to-end latency constraints (to ensure timeliness of results), and exhibit a great amount of degeneracy. A given processor can typically work on any of several computational workflows. Similarly, a workflow can be performed by (a sequence of) any of multiple processors. Moreover, failures that delay one workflow (such as anomalous conditions that lead the workflow to consume an unnaturally large amount of time and resources) can also delay other workflows that compete on the same resources. Recent work addressed greedy heuristics for allocating resources to real-time flows in such a way that resiliency of end-to-end flow latency is maximized with respect to local failures [15]. It was shown that the end-to-end latency can be made significantly more resilient to perturbations by a proper resource reallocation that minimizes dependencies between flows that arise from resource sharing. A broader look at resiliency to function-loss under a more general failure model and time constraints remains a matter of future investigation.

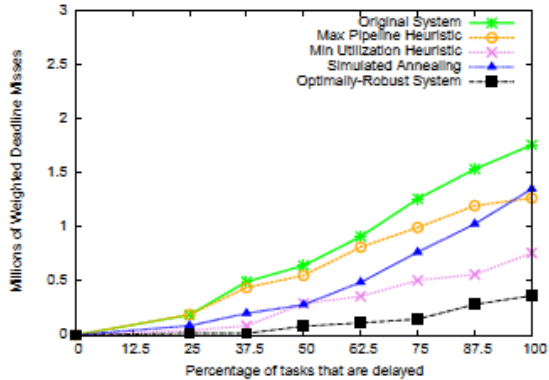

**Figure 5.** Resiliency of end-to-end flow latency can be maximized with respect to local delays by properly allocating resources to flows. (Adapted from [15])

### 7. Resiliency to Input Corruption

In multi-genre networks that span physical, information, and social spaces, correctness of network functions and timing, addressed above, is not in itself sufficient to guarantee resiliency. Network functions only transform inputs to outputs. The correctness of outputs depends not only on the correctness of the transformation, but also on the integrity of inputs. Hence, a functionally-resilient network may remain vulnerable to input corruption. The problem is especially severe in networked information and social systems. Such networks operate in the presence of high uncertainty (as in a battlefield or a disaster-response scenario). Techniques are needed to ensure resiliency of the quality of information delivered.

An "optimal" way of ensuring quality of information might trivially be to use highly reliable sources. This approach, even where feasible, is vulnerable to failure or corruption of those sources. Therefore, of much interest is the development of mechanisms that maintain high quality of information in a manner resilient to individual source failure or corruption. Hence, while significant literature exists on various data fusion and outlier elimination methods [16], we focus on techniques where we do not rely on any one source for correctness of information, but rather infer both the quality of information and the credibility of sources directly from the data itself.

The core of these resiliency mechanisms (to input corruption) lies in exploiting the topology of linkages that exist between data inputs to rank information dynamically in a way that allows the system to adapt to increasing amounts of noise and bad data by identifying and removing such data (and its sources) from the input pool. A static version of this idea takes root from Google's PageRank [17]. PageRank infers the most credible sources of information for a given query, with no *a priori* knowledge of the authority of each source. It does so by recursively observing a network of links between pages, where more links to a page (from more credible sources) imply more credibility to the page, leading to an iterative assessment (or ranking) of credibility of all pages. The technique can be generalized to the joint estimation of credibility of arbitrary *sources* and *claims*. Given a collection of sources and a collection of claims, the credibility of claims is a function of the number and credibility of their sources, whereas the credibility of sources is a function of the credibility of their claims. Several prior heuristics exist for expressing this recursive relation with the purpose of simultaneously arriving at a ranking of sources and a ranking of claims by credibility, given the reported data itself whose credibility is being determined (e.g., [18,19,20]). Most recently, an iterative function was derived that was shown to be *optimal* in that it converges to the maximum likelihood estimate for the probability of correctness of sources and claims. Furthermore, a bound was derived on the confidence interval of this maximum likelihood estimate [21]. Applying these results in real-time to an incoming data stream allows for continual assessment of input quality. Should quality of some inputs be suddenly degraded, the mechanism soon catches up and downgrades the credibility ranking of the inputs in question. This adaptation offers resiliency to input contamination.

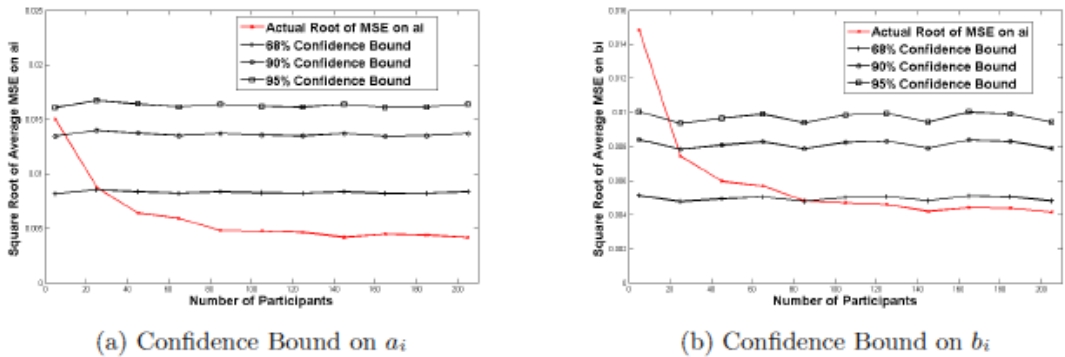

(a) Confidence Bound on $a_i$     (b) Confidence Bound on $b_i$

**Figure 6.** Confidence in correctness of sources (described by parameters $a_i$ and $b_i$) can be managed with resiliency within a desired bound. (Adapted from [21].)

The approach can easily accommodate additional prior knowledge, when available, but does not have to depend on such knowledge. The main advantage of this recent result is that it offers not only the best (i.e., maximum likelihood) hypothesis regarding the probability of correctness of data and sources, but also a confidence value in this hypothesis. Quantifying the degree to which the confidence interval (and hence quality of information) is susceptible to failures in individual sources (as a function of the topology of the source-claim network) remains an interesting open problem.

## 8. Resiliency of Inference

The last element common to multi-genre networks lies in the algorithms that extract information from data. The ultimate success of these networks often lies in delivering high-quality, high-confidence support for decision making. A related question becomes: how much resiliency can be built into the decision support algorithms themselves, such that they may gracefully adapt to failures of the underlying networks to deliver sufficient data, failures to deliver it at sufficient quality, or failures to deliver it at an affordable cost?

The goal is to make sure that high quality outputs can be restored even as the amount and quality of available resources and data decrease. This is achieved by finding "adaptation knobs" that restore acceptable performance by reallocating scarce resources to where they are needed most for the situation at hand. Applying this wisdom to the design space of algorithms (e.g., algorithm that make predictions from current observations), one may observe that by breaking up the data space appropriately into the right sub-cases, one is able to construct specialized algorithms for each sub-case. Being specialized, these algorithms are likely to contain much fewer parameters, which decreases their cost. Moreover, being specialized, they can also do a better job at estimating or predicting system behavior correctly in the special case to which they pertain. The combination of decreased cost and increased accuracy offers the desired improvement in quality/cost ratio. The underlying adaptation capability lies in recognizing which algorithm to use in a particular scenario. Consistent with the above ideas, recent work presented sparse regression cubes [22], an approach borrowed from data mining that jointly (i) partitions sparse, high-dimensional data into subspaces within which reliable prediction models apply and (ii) determines the best such model for each partition. It was shown that by adapting the model to the current situation, as indicated by the structure of the sparse regression cube, significantly more resilient predictions can be made in the face of uncertainty, errors, and resource constraints.

With different resiliency mechanisms explored at different layers of the multi-genre network, a significant research question that remains is to understand the combined behavior of such a heterogeneous network ecosystem, given our understanding of the behavior of its individual components. This question has been a key motivation and grand challenge of network science; the science of prediction, measurement, and adaptation of complex multi-genre networks. While this paper touched on some of its key results, fundamental discoveries are yet to be made in understanding performance, resiliency, and adaptation properties of complex systems operating in physical, information, and social spaces.

**References**

[1] M. E. J. Newman, S. H. Strogatz, and D. J. Watts, "Random graphs with arbitrary degree distributions and their applications," Physical Review E, Volume 64, No. 2, 2001

[2] M. E. J. Newman, Michelle Girvan, J. Doyne Farmer, "Optimal design, robustness, and risk aversion," Physical Review Letters, Vol. 89, No. 2, July 2002

[3] R. Albert, H. Jeong, A.-L. Barabási, "Error and attack tolerance of complex networks," Nature, pp. 378–482, 2000

[4] Duncan S. Callaway, M. E. J. Newman, Steven H. Strogatz, and Duncan J. Watts, "Network Robustness and Fragility: Percolation on Random Graphs," Physical Review Letters, Vol. 85, No. 25, December 2000

[5] L. Sha, "Using Simplicity to Control Complexity," IEEE Software, July-August, 2001

[6] Sergey V. Buldyrev, Roni Parshani, Gerald Paul, H. Eugene Stanley, Shlomo Havlin, "Catastrophic cascade of failures in interdependent networks," Nature Letters, April 2010

[7] Larry Blume, David Easley, Jon Kleinberg, Robert Kleinberg, Eva Tardos, "Which Networks Are Least Susceptible to Cascading Failures?" In Proc. 52nd Annual IEEE Symposium on Foundations of Computer Science, Palm Springs, California, October 2011

[8] Duncan Watts, "A Simple Model of Global Cascades on Random Networks," Proceedings of the National Academy of Sciences, Vol. 99, No. 9, April 2002

[9] James M Whitacre and Axel Bender, "Networked buffering: a basic mechanism for distributed robustness in complex adaptive systems," Theoretical Biology and Medical Modeling, Vol. 7, No. 20, June 2010

[10] Frei R., Whitacre J.M. "Degeneracy and Networked Buffering: principles for supporting emergent evolvability in agile manufacturing systems" Journal of Natural Computing, Special Issue on Engineering Emergence, *in press*

[11] Alvin C. Valera, Winston K.G. Seah, and S.V. Rao, "Improving Protocol Robustness in Ad Hoc Networks through Cooperative Packet Caching and Shortest Multipath Routing," IEEE Transactions on Mobile Computing, Vol. 4, No. 5, September 2005

[12] Adilson E. Motter and Ying-Cheng Lai, "Cascade-based attacks on complex networks," Physical Review E, Vol. 66, 2002.

[13] G. Korniss, R. Huang, S. Sreenivasan, and B.K. Szymanski, "Optimizing Synchronization, Flow, and Robustness in Weighted Complex Networks," Handbook of Optimization in Complex Networks, edited by My T. Thai and P. Pardalos, Springer, 2011, *in press*

[14] W. Gao, G. Cao, A. Iyengar, and M. Srivatsa, "Supporting Cooperative Caching in Disruption Tolerant Networks," IEEE International Conference on Distributed Computing Systems (ICDCS), 2011

[15] Praveen Jayachandran and Tarek Abdelzaher, "On Structural Robustness of Distributed Real-Time Systems Towards Uncertainties in Service Times," in Proc. IEEE Real-Time Systems Symposium (RTSS), December 2010

[16] Martin E. Liggins, David L. Hall, James Llinas, "Handbook of Multisensor Data Fusion," CRC Press, 2009

[17] S. Brin and L. Page, "The anatomy of a large-scale hypertextual web search engine," In 7th international conference on World Wide Web (WWW'07), pp. 107–117, 1998

[18] J. M. Kleinberg, "Authoritative sources in a hyperlinked environment," Journal of the ACM, Vol. 46, No. 5, pp. 604–632, 1999

[19] J. Pasternack and D. Roth, "Knowing what to believe (when you already know something)," In Proc. International Conference on Computational Linguistics (COLING), 2010

[20] X. Yin, J. Han, and P. S. Yu, "Truth discovery with multiple conflicting information providers on the web," IEEE Transactions on Knowledge and Data Engineering, Vol. 20, pp. 796–808, June 2008

[21] Dong Wang, Tarek Abdelzaher, Lance Kaplan, Charu Aggarwal, "On Quantifying the Accuracy of Maximum Likelihood Estimation of Participant Reliability in Social Sensing," In Proc. 8th International Workshop on Data Management for Sensor Networks, Seattle, WA, August 2011

[22] Hossein Ahmadi, Tarek Abdelzaher, Jiawei Han, Raghu Ganti and Nam Pham, "On Reliable Modeling of Open Cyber-physical Systems and its Application to Green Transportation," ICCPS, Chicago, IL, April 2011

[23] John C. Doyle, David L. Alderson, Lun Li, Steven Low, Matthew Roughan, Stanislav Shalunov, Reiko Tanaka and Walter Willinger, "The Robust Yet Fragile Nature of the Internet," Proceedings of the National Academy of Sciences of the United States of America, 2005

[24] Zhaojun Bai, Patrick M. Dewilde, and Roland W. Freund, "Reduced-Order Modeling," Numerical Methods in Electromagnetics, of Numerical Analysis, Vol. XIII, Amsterdam: Elsevier 2005, pp. 825-895

[25] Chandrasekharan, P., C., "Robust Control of Linear Dynamical Systems," Academic Press, 1996

[26] Walker, B., Holling, C. S., Carpenter, S. R., Kinzig, A. "Resiliency, adaptability and transformability in social–ecological systems," Ecology and Society, Vol. 9, No. 2, 2004, http://www.ecologyandsociety.org/vol9/iss2/art5

[27] Gunderson, L.H., "Ecological Resiliency — In Theory and Application," Annual Review of Ecology & Systematics 31: 425, 2000

[28] C. Perrow, "Normal Accidents: Living with High Risk Technologies," Princeton University Press, 1984

[29] Alexander Kott, "Battle of Cognition: The Future Information-Rich Warfare and the Mind of the Commander", *Greenwood Publishing Group*, 2008, pp. 205-211

[30] Alexander Kott, "Information Warfare and Organizational Decision-making" *Artech House*, 2006.

[31] Alexander Kott and Gary Citrenbaum "Estimating Impact: A Handbook of Computational Methods and Models for Anticipating Economic, Social, Political and Security Effects in International Interventions" Springer, 2010, pp 2-14

[32] Network Science Collaborative Technology Alliance (www.ns-cta.org)

[33] Fiona Chandra, Dennice F. Gayme, Lijun Chen, John C. Doyle, "Robustness, Optimization, and Architectures," *European Journal of Control*, 2011, pp. 472–482